\documentstyle[12pt]{article}
\def\bfone{\relax{\rm 1\kern-.35em 1}}
\def\IR{\relax{\rm I\kern-.18em R}}
\font\cmss=cmss10 \font\cmsss=cmss10 at 7pt
\def\ZZ{\relax\ifmmode\mathchoice
{\hbox{\cmss Z\kern-.4em Z}}{\hbox{\cmss Z\kern-.4em Z}}
{\lower.9pt\hbox{\cmsss Z\kern-.4em Z}}
{\lower1.2pt\hbox{\cmsss Z\kern-.4em Z}}\else{\cmss Z\kern-.4em
Z}\fi}
\def\ee#1{{\rm e}^{#1}}
\def\ii{{\rm i}}
\def\twomat#1#2#3#4{\left(\begin{array}{cc}#1& #2\\ #3 &
#4\end{array}\right)}

\begin{document}
\vspace*{3.5 cm}
\begin{center}
\hfill POLFIS-TH/95-12
\vskip.4in
{\large\bf R-Symmetry, twisted $N=2$ Theory and the Role of the Dilaton }
\end{center}
\vspace*{6.0 ex}
\begin{center}
{\large\rm Riccardo D'Auria}
\end{center}
\vspace*{1.5 ex}
\begin{center}
{\large\it Dipartimento di Fisica, Politecnico di Torino\\
Corso Duca Degli Abruzzi 24, I-10129 Torino, Italy}
\end{center}
\vspace*{4.5 ex}
\begin{center}
{\bf Abstract}
\end{center}
\vspace*{3.0 ex}

{\small We discuss R-symmetry in locally supersymmetric $N=2$ gauge theories
coupled to hypermultiplets, which can be viewed as effective theories of
heterotic string models. In this type of supergravities a suitable
R-symmetry exists and can be used to topologically twist the theory.
The vector multiplet of the dilaton-axion field has a different R-charge
assignment with respect to the other vector multiplets.}

\vspace{3.0 ex}

In his first paper on topological field theories
\cite{topfgen_7}, Witten had shown how to derive a topological
reinterpretation of N=2 Yang--Mills theory in four--dimensions by
redefining the Euclidean Lorentz group:
\begin{equation}
SO(4)_{spin} \, =\, SU(2)_L \, \otimes \, SU(2)_R
\label{spingroup}
\end{equation}
in the following way:
\begin{equation}
SO(4)_{spin}^{\prime} \, =\, SU(2)_L \, \otimes \, SU(2)_R^\prime
\ ;\qquad SU(2)_R^\prime \, = \, {\rm diag} \left ( SU(2)_I \,
\otimes \, SU(2)_R \right )\ .
\label{newspingroup}
\end{equation}
Here $SU(2)_I$ is a factor of the automorphism group
 $SU_I(2)\times U_I(1)$  of the  N=2 supersymmetry algebra.
Moreover the second factor $U_I(1)$, usually called R-symmetry group,
enters in a suitable redefinition of the $U_g(1)$ ghost number group:
\begin{equation}
U_g(1)\to U'_g(1)={\rm diag} \left( U_g(1)\times U_I(1)\right)
\end{equation}
According to this prescription
the quantum numbers of a generic field change as follows
\begin{equation}
\ ^r (L,R,I)^g \to (L,R\times I)^{g+r}
\ ,\end{equation}
where $L,R,I,R\times I$ label the representations of the original three
$SU(2)$ defined in equations (\ref{spingroup}), (\ref{newspingroup}), and
$r,g$ are the charges of the $U_I(1)$ and $U_g(1)$ groups respectively.

In particular, for physical fields ($g=0$), the R-symmetry charge $r$ coincides
with the ghost number of the topologically twisted theory. This procedure
is generally referred to as topological twisting.
This twisting procedure applied to the
rigid N=2 supersymmetric Yang--Mills theory
 yields topological Yang--Mills theory, where the fields of the
 N=2 supermultiplet have the following reinterpretation:
 \begin{eqnarray}
 \mbox{gauge~boson} ~~A_\mu^\alpha &\rightarrow & {\rm phys.~field}
 \quad q_{gh}\, =
 \, 0 \nonumber\\
\mbox{left--handed gaugino} ~~\lambda^{\alpha A} &\rightarrow &
{\rm top.~ghost} \quad q_{gh}\, =
 \, 1 \nonumber\\
\mbox{right--handed gaugino} ~~\lambda_{\alpha^\star}^{A} &\rightarrow &
\mbox{top.~antighost} \quad q_{gh}\, =
 \, -1 \nonumber\\
\mbox{scalar} ~~Y^{I} &\rightarrow &
\mbox{ghost~for~ghosts} \quad q_{gh}\, =
 \, 2 \nonumber\\
\mbox{conjug.~scalar} ~~{\bar Y}^{I^\star} &\rightarrow &
\mbox{antighost~for~antighosts} \quad q_{gh}\, =
 \, -2
\label{ghostnumeri}
\end{eqnarray}
Actually, this is the result one obtains starting from the so called
minimal coupling of the $N=2$ rigid theory, that is when the prepotential
$F(Y)$ is quadratic in the scalar fields $Y^I$.

In order to extend Witten's ideas to the case of an arbitrary
N=2 theory including gravity and hypermultiplets, four steps,
that were clarified in refs.
\cite{topftwist_1,topftwist_2}, are needed:
\begin{enumerate}
\item[i)] Systematic use of the BRST quantization, prior to the twist.
\item[ii)]
Identification of a gravitationally extended R-symmetry
that can be utilized to redefine the ghost--number in the
topological twist.
\item[iii)] In presence of hypermultiplets, a
further modification of rule (\ref{newspingroup}) for the
redefinition of the Lorentz group that becomes:
\begin{equation}
SO(4)_{spin}^{\prime} \, =\, SU(2)_L^\prime \, \otimes \, SU(2)_R^\prime
\quad \quad \cases{
SU(2)_R^\prime \, = \, {\rm diag} \left ( SU(2)_I \,
\otimes \, SU(2)_R \right )\cr
SU(2)_L^\prime \, = \, {\rm diag} \left ( SU(2)_Q \,
\otimes \, SU(2)_L \right )\cr }
\label{newnewspingroup}
\end{equation}
Here $SU(2)_Q$ is a group whose action vanishes on all fields
except on those of the hypermultiplet sector, so that its role was not
perceived in Witten's original case.
\item[iv)] Redefinition of the supersymmetry ghost field {(\it topological
shift)}.
\end{enumerate}
Points { i)} and { iv)} of the above list do not impose
any restriction on the scalar
manifold geometry, so we do not discuss them further.

It is clear that,
when N=2 supersymmetry is made local, $R$-symmetry should extend to
a suitable symmetry of matter coupled supergravity.
Indeed, the possibility of topological twisting in the gravitational case
requires a well defined R-symmetry charge for the parent $N=2$ local theory
in order that the new ghost numbers $g'=g+r$ be well defined for all the
fields. Thus, the problem of the identification of the R-symmetry in $N=2$
local
theory and the possibility of defining a suitable twisting for topological
gravity amount to the same thing.
It turns out \cite{rsym}
that in order to have a consistent definition of R-symmetry in the
local case,
an essential ingredient is the presence of
an axion-dilaton vector multiplet (v.m.) in the matter coupled theory which
behaves in the reversed manner as far as the R-symmetry charge
 (and hence ghost number)
assignments are concerned, namely:

\begin{eqnarray}
 \mbox{gauge~boson} ~~A_\mu^s &\rightarrow & {\rm ghost.~for~ghost}
 \quad q_{gh}\, =
 \, 2 \nonumber\\
\mbox{left--handed gaugino} ~~\lambda^{s A} &\rightarrow &
{\rm top.~antighost} \quad q_{gh}\, =
 \, -1 \nonumber\\
\mbox{right--handed gaugino} ~~\lambda_{s^\star}^{A} &\rightarrow &
\mbox{top.~ghost} \quad q_{gh}\, =
 \, 1 \nonumber\\
\mbox{scalar} ~~S &\rightarrow &
\mbox{\rm phys.~field} \quad q_{gh}\, =
 \, 0 \nonumber\\
\mbox{conjug.~scalar} ~~{\bar S} &\rightarrow &
\mbox{\rm phys.~field} \quad q_{gh}\, =
 \, 0 .
\label{ghostnumeriprimo}
\end{eqnarray}
We note now that an axion-dilaton v.m. behaving in a different way with
respect to R-symmetry transformations is what naturally shows up in
the low energy limit of the heterotic string compactified on  six
dimensional manifolds, like
 $T_2\times K3$\cite{skgsugra_16,ferpor,ferkouncos}.
In this context, at the classical level
 the following vector multiplet and hypermultiplet scalar manifold $ST(n)$ and
$HQ(m)$ respectively are given by
\begin{equation}
\label{abc}
 ST(n) {\stackrel{\rm def}{=}} {{SU(1,1)}\over
{U(1)}}\, \otimes {{SO(2,n)}\over{SO(2)\otimes SO(n)}}\ \ \ ,\ \
 HQ(m ) {\stackrel{\rm def}{=}}
{{SO(4,m)}\over{SO(4)\otimes SO(m)}}\ .
\end{equation}
Here $n,m$ are the numbers of vector multiplets and hyper\-multiplets
re\-spectively
and $SU(1,1)/U(1)$ is parametrized by the dilaton complex field $S$.
We will see below that the R-symmetry assignments of eq.
(\ref{ghostnumeriprimo}) can be easily computed in this case.
Furthermore, the very presence of the axion-dilaton v.m. in the
$N=2$ local theory solves an apparent difficulty which was noted in
ref. \cite{topftwist_1,topftwist_2}
in a different approach to the problem of the R-charges
assignments for the gravitationally coupled $N=2$ theory. There, it
appeared difficult to saturate the sum rule
\begin{equation}
\sum_{i=1}^{n} (gh)^i=3\tau
\end{equation}
relating the ghost numbers $(gh)^i$ of the various fields, typically an even
number, to the dimensionality $3\tau$ of the moduli space of a typical
gravitational instanton, like an ALE manifold (here $\tau$ is the Hirzebruch
signature). However, the presence of the axion-dilaton v.m. allows the
complexification of the moduli space for the K\"ahler structure deformations
enlarging the dimensionality of the moduli space to $4\tau$, which, being
an even number, solves the aforementioned problem.

Coming back to our main problem, we recall
that while in the classical case the R-symmetry is a $U(1)$
group, it breaks down to $\ZZ_p\in U(1)$ when non-perturbative quantum
corrections are turned on.

Furthermore it happens that in the classical $ST(n)$
case the continuous R--symmetry is an off--shell symmetry of the
action while in the quantum ${\hat {ST}}(n)$ case the
discrete R--symmetry acts in general as an electric--magnetic
duality rotation of the type of S--duality.
As stated before, the $R$-symmetry of
rigid N=2 gauge theories should have a natural extension
to the gravitationally
coupled case.
We need some guidelines
to relate the R-symmetry of a rigid theory to the R-symmetry
of  a corresponding locally supersymmetric theory. The main points
to have in mind are the following ones:
\begin{itemize}
\item{The R-symmetry group $G_R$, whether continuous or discrete,
must act on the
symplectic sections $( X^\Lambda, F_\Lambda )$ ($\Lambda=0,\ldots,n$)
 by means of symplectic
matrices:
\begin{equation}
\label{nuova}
\begin{array}{ccc}
\forall \, g \, \in \, G_R & \hookrightarrow & \left (\matrix{A(g) &
B(g) \cr C(g) & D(g) \cr } \right ) \, \in \, \Gamma_R \subset
SP(4+2n,\IR).
\end{array}
\end{equation} }
\item{The fields of the theory must have under $G_R$ well defined
 charges, so that
$G_R$ is either a $U_R(1)$ group if continuous or a cyclic group
$\ZZ_p$ if discrete.}
\item{By definition the left--handed and right--handed gravitinos must
have R--charges $q=\pm 1$, respectively}
\item{Under the $G_R$ action there must be, in the special manifold, a
preferred direction corresponding to the dilaton--axion multiplet
whose
R--charges are reversed with respect to those of all the other
multiplets. As emphasized, this is necessary, in order
for the topological twist to leave the axion--dilaton
field physical in the topological theory, contrary to
the other scalar partners of the vectors that become
ghosts for the ghosts}
\end{itemize}
\par
As we will discuss below,
for the classical coset manifolds
$ST(n)$  the appropriate
R-symmetry is continuous and it is easily
singled out: it is the
 $SO(2) \sim U(1)$ subgroup
of the isotropy group $SO(2) \times SO(n) \, \subset \, SO(2,n)$.
The coordinates that diagonalize the R--charges
are precisely the Calabi--Visentini coordinates of the classical
theory\cite{rsym}. In the flat limit they can be identified  with
the special coordinates of rigid special geometry. Hence such
gravitational R-symmetry is, as required, the supergravity
counterpart  of the R-symmetry considered in the rigid
theories.  Due to the direct product structure of
this classical manifold
the  preferred direction corresponding to the dilaton--axion field
 is explicitly singled out in the
$SU(1,1)/U(1)$ factor .
\par
Generically,
in the quantum case, the R-symmetry group $G_R$ is discrete.
Its action on the  quantum counterpart of the Calabi--Visentini
coordinates ${\hat Y}^\alpha$ must approach the action of a discrete
subgroup of the classical $U(1)_R$ in the same asymptotic region
where the local geometry of the quantum manifold ${\hat {ST}}(n)$
approaches that of $ST(n)$. This is  the large radius limit if we
think of
${\hat {ST}}(n)$ as of the moduli--space of some dynamical
Calabi--Yau threefold. To this effect recall that special K\"ahler
geometry is the moduli--space geometry of Calabi--Yau threefolds
and we can generically assume that any special manifold ${\cal SM}$
corresponds to some suitable threefold. Although the $G_R$
group is, in this sense, a subgroup of the classical $U_R(1)$
group, yet we should not expect that it is realized by a subgroup
of the symplectic matrices that realize $U(1)_R$ in the classical
case.  The different structure of the symplectic R--matrices is
precisely what allows a dramatically different form of the
special metric in the quantum and  classical case.
The need for this difference can be perceived {\it a priori} from
the request that the quantum R-symmetry matrix should be
symplectic integer valued. As we are going to see this is possible
only for $\ZZ_4$ subgroups of $U(1)_R$ in the original
symplectic embedding. Hence  the different
$\ZZ_p$ R--symmetries appearing  in rigid  quantum  theories
should have different  symplectic embeddings in the gravitational case.

\par
Let us now give the general properties of the gravitationally extended
 R-symmetry.
R-symmetry is either a $U(1)$ symmetry or
a discrete $\ZZ_{p}$ symmetry.
Thus, if $R$-symmetry acts diagonally with charge $q_R$ on a field $\phi$,
this means  that $\phi\rightarrow \ee{q_R\ii \vartheta}
\phi$, $\vartheta\in [0,2\pi]$ in the continuous case.
In the discrete case only the values
$\vartheta=\frac{2\pi}{p}l$,  $l=0,1\ldots p-1$ are allowed
and in particular
the generator of the $\ZZ_{p}$ group acts as  $\phi\rightarrow
 R\phi =
\ee{q_R\frac{2\pi\ii}{p}}\phi$.
\par
By definition $R$-symmetry acts diagonally with charge $+1$ ($-1$)
on the left-(right)-handed gravitinos (in the same way as
it acts on the supersymmetry parameters in the rigid case):
\begin{equation}
\label{so2n11}
\begin{array}{c}
\psi_A\rightarrow\ee{\ii\vartheta}\psi_A\\
\psi^A \rightarrow\ee{-\ii\vartheta}\psi^A
\end{array}
\hskip 1cm \mbox{i.e.}\hskip 1cm
\begin{array}{l}
q_L(\psi_A) = 1\\
q_R(\psi^A) = -1 .
\end{array}
\end{equation}
$R$-symmetry
generates isometries  $z^i\rightarrow (R_{2\vartheta}z)^i$
of the scalar metric  $g_{ij^*}$ and it is embedded into
$Sp(2n+4,\IR)$ by means of  a symplectic matrix:
\begin{equation}
\label{Mmatrix}
 M_{2\vartheta} =
\left(\begin{array}{cc}  a_{2\vartheta} & b_{2\vartheta}
\\c_{2\vartheta} &
d_{2\vartheta}
\end{array}\right)   \, \in \, Sp(2n+4,\IR).
\end{equation}
\par\noindent
As we have already pointed out it turns out that
in the classical case of $ST(n)$ manifolds R--symmetry
does not mix the Bianchi identities with the field equations
since the matrix (\ref{nuova}) happens to be block diagonal:
$b_{2\theta}=c_{2\theta}$.
In the quantum case, instead, this is in general not true.
There is a symplectic action on the section
 $(X^\Lambda,F_\Lambda)$
induced by $z^i\rightarrow (R_{2\vartheta}z)^i$:
\begin{equation}
\label{fattorello}
(X,F)\rightarrow f_{2\vartheta}(z^i)  {M}_{2\vartheta} \cdot (X,F)
\end{equation}
where the K\"ahler compensating factor $f_{2\vartheta}(z^i)$ depends in
general both on the transformation parameter $\vartheta$ and on the
base--point $z$. By definition this compensating factor is the same
that appears in the transformation of the gravitino field $\psi_A \,
\to \, \exp [f_{2\vartheta}(z^i)/2] \, \psi_A$. Since  we have
imposed  that the transformation of the gravitino field should be as
in (\ref{so2n11}) it follows that the R-symmetry transformation
must be such as to satisfy eq.(\ref{fattorello}) with a suitable matrix
(\ref{Mmatrix}) and with a compensating K\"ahler factor
of the following specific form:
\begin{equation}
\label{ftheta}
f_{2\vartheta}(z^i) = \ee{2\ii\vartheta} .
\end{equation}
Condition $(\ref{ftheta})$ is a crucial constraint on the form of
$R$-symmetry.\par\noindent
Given these inputs, it turns out that the susy transformation laws of
the $N=2$ local theory\cite{skgsugra_1}
 fix in a unique way the R-symmetry assignments for
all the fields, and in particular give the charge assignment of eq.
(\ref{ghostnumeriprimo})
of the axion-dilaton multiplet.

In the case of the microscopic classical $N=2$ lagrangian,
the special K\"ahler manifold of the v.m.  scalars is a
$ST(n)$ manifold.
Consider the following parametrization of the $N=2$ symplectic sections
\cite{ceretoine}
\begin{equation}
 \label{so2nssec}
(X^{\Lambda}, F_{\Lambda})=(X^{\Lambda}, S \eta_{\Lambda\Sigma}
X^{\Sigma})\ \ \ ,\ \   X^{\Lambda} =
\left(\begin{array}{c} 1/2\hskip 2pt (1 + Y^2) \\ \ii/2\hskip 2pt (1 -
Y^2)\\ Y^{\alpha}\end{array}\right).
\end{equation}
In eq.
(\ref{so2nssec}) $Y^\alpha$ are
the Calabi--Visentini coordinates,
parametrizing the coset manifold \linebreak
$SO(2,n)/SO(2)\times SO(n)$, while $S$ parametrizes $SU(1,1)/U(1)$. The
pseudoorthogonal metric $\eta_{\Lambda\Sigma}$
has the signature $(+, +,
-, \ldots, -)$.
In this parametrization the action of the isometry group
$SU(1,1)\times SO(2,n)$ is linearly realized, and
the action of  $R$-symmetry is extremely simple.
One finds
\begin{equation}
\label{sknrsim1}
 S \rightarrow S\ \ \ ,\ \  Y^\alpha \rightarrow
\ee{2\ii\vartheta} Y^\alpha
\end{equation}
according to eq. (\ref{ghostnumeriprimo}).

Utilizing the explicit form eq. (\ref{so2nssec}) of the symplectic section,
eq. (\ref{sknrsim1}) induces the transformation:
\begin{equation}
\label{sknrsim2}
\begin{array}{l}
\left(\begin{array}{c}X\\F\end{array}\right) \rightarrow \ee{2\ii\vartheta}
\left(\begin{array}{cc}{ m}_{2\vartheta} & 0\\ 0 &
({ m}^T_{2 \vartheta} )^{-1}\end{array}\right)
\left(\begin{array}{c}X\\F\end{array}\right)\\ \null \\
{ m}_{2\vartheta} =
\left(\begin{array}{ccc}\cos 2\vartheta & -\sin 2\vartheta & 0\\
\sin 2\vartheta & \cos 2\vartheta & 0\\
0 & 0 & \bfone_{n\times n}\end{array}\right)\in SO(2,n) .\end{array}
\end{equation}
We see that the crucial condition (\ref{ftheta}) is met. Furthermore note
that in this classical case $b_{2\vartheta}=c_{2\vartheta}=0$,
the matrix (\ref{Mmatrix}) is completely diagonal
and it has the required eigenvalues $(\ee{\ii\theta},\ee{-\ii\theta},
1,\ldots,1)$.\par
At this point we need no more checks; the $R$-symmetry defined by
eq. (\ref{sknrsim1}) is a true symmetry of the lagrangian and satisfies
all the expected properties.
The gauge fields $A^\alpha$ do not transform, while the $A^0,A^S$ gauge
fields undergo an $SO(2)$ rotation:
\begin{equation}
\label{sknrsim6}
\left(\begin{array}{c}A^0\\ A^S\end{array}\right)  \rightarrow
\twomat{\cos 2\vartheta}{-\sin 2 \vartheta}{sin 2\vartheta}{\cos 2
\vartheta} \left(\begin{array}{c}A^0\\ A^S\end{array}\right) \ \ ,\ \
A^\alpha  \rightarrow A^\alpha \ .
\end{equation}
Notice that from eq.s (\ref{sknrsim2}), (\ref{sknrsim6})
and from the explicit form of the
embedding\cite{rsym},
we easily check that the R-symmetry for the $ST(n)$
case is nothing else but the $SO(2)\sim U(1)$ subgroup
of the isometries appearing in the denominator of the
coset $SO(2,n)/SO(2)\times SO(n)$.
\par
At the quantum level the $R$--symmetries should act
on the symplectic sections as a symplectic matrix  belonging
to $Sp(2n + 4, \ZZ)$.  Consider then the intersection
of the continuous $R$ symmetry of eq.s (\ref{sknrsim1},\ref{sknrsim2})
with  $Sp(2n + 4, \ZZ)$: the result is a $\ZZ_4$ R-symmetry
generated by the matrix  $M_{2\vartheta}$ with $\vartheta=\pi /4$,
where:
\begin{equation}
\label{discreta}
m_{\pi /2}=\left(\begin{array}{ccc}0 & -1 & 0\\
1 & 0 & 0\\
0 & 0 & \bfone_{n\times n}\end{array}\right)\in SO(2,n; \ZZ) .
\end{equation}
As already observed, in a generic case, after the quantum
corrections are implemented, the discrete R-symmetry
$\ZZ_p$ is a subgroup of $U(1)_R$ as far as the action
on the moduli at large values is concerned, but it is
implemented by $Sp(4+2n,\ZZ)$ matrices that are
not the restriction to discrete value of theta of the
matrix $M_{2\vartheta}$ defined in eq.s (\ref{sknrsim2}).
In the one modulus case where,  according to the
analysis by Seiberg--Witten the rigid R-symmetry is $\ZZ_4$,
there is the possibility of maintaining the classical form of
the matrix $M_{2\vartheta}$ also at the quantum level and
in the case of local supersymmetry. This seems to be
a peculiarity of the one--modulus N=2 gauge theory.
\par

In the classical case of the $ST(n)$ manifolds the existence of
a preferred direction is obvious from the definition of the
manifolds and R--symmetry singles it out in the way discussed.
Let us see how the dilaton--axion
direction can be singled out by the discrete R--symmetry
of the quantum manifolds ${\hat {ST}}(n)$.  Let $G_R=\ZZ_p$
and let $\alpha=e^{2\pi{\rm i}/p}$ be a $p$--th root of the
unity.  In the space of the scalar fields $z^{i}$ there always
will be a coordinate basis $\{ u^{i} \}$ $(i=1,\dots \, n+1)$
that diagonalizes the action of $R_{2\vartheta}$ so that:
\begin{equation}
R_{2\vartheta} \, u^{i} ~=~\alpha^{ q_i} \, u^{i} \quad \quad
q_i = 0,1, \dots , p-1 \, \mbox{mod} \, p
\label{drago_1}
\end{equation}
The $n+1$ integers $q_i$ (defined modulo $p$) are the
R--symmetry charges of the scalar fields $u_i$.
At the same time a generic $Sp(4+2n,\IR)$ matrix
has eigenvalues:
\begin{equation}
\left ( \lambda_0, \lambda_1, \dots , \lambda_{n+1}, {{1}\over{\lambda_0}},
{{1}\over{\lambda_1}},
\dots , {{1}\over{\lambda_{n+1}}} \right )
\label{autovalores}
\end{equation}
The R--symmetry symplectic matrix $M_{2\vartheta}$ of
eq. (\ref{Mmatrix}), being the generator of a cyclic group
$\ZZ_p$, has eigenvalues:
\begin{equation}
\lambda_0 = \alpha^{k_0}, \quad \lambda_1 = \alpha^{k_1}, \quad
\dots ,\ \lambda_{n+1} = \alpha^{k_{n+1}}\ ,
\label{specialmente}
\end{equation}
where $(k_0 , k_1 , \dots \, k_{n+1} )$ is a new set of
$n+2$ integers defined modulo $p$. These numbers are
the R--symmetry charges of the electric--magnetic field strenghts
\begin{equation}
F_{\mu\nu}^0+{\rm i} \, G_{\mu\nu}^0, \quad F_{\mu\nu}^1
+ {\rm i} \, G_{\mu\nu}^1 ,\quad \dots \,\quad F_{\mu\nu}^{n+1}
+ {\rm i} \, G_{\mu\nu}^{n+1}\ ,
\end{equation}
their negatives, as
follows eq. (\ref{autovalores}), being the charges
of the complex conjugate combinations
$F_{\mu\nu} - {\rm i} G_{\mu\nu}$.
Since what is really
relevant in the topological twist are the differences of
ghost numbers (not their absolute values), the interpretation
of the scalars $u^{i}$ $(i=1, \dots , n)$ as ghost for ghosts and of
the corresponding vector fields as physical gauge fields
requires that
\begin{equation}
q_i ~=~k_i \, + \, 2 \quad\quad  i=1, \dots , n
\label{shifto}
\end{equation}
On the other hand, if the vector partner of the
axion--dilaton field has to be a ghost for ghosts,
the $S$--field itself being physical, we must
have:
\begin{equation}
k_{n+1} = q_{n+1} + 2
\label{inversamente}
\end{equation}
In eq. (\ref{inversamente}) we have conventionally identified
\begin{equation}
S=u^{n+1}
\end{equation}
Finally the R--symmetry charge $k_0$ of the last vector field--strength
$F^0_{\mu\nu}$
is determined by the transformation law
of the graviphoton combination \cite{rsym}.

\par
In \cite{monodrcy} an explicit example is
provided of quantum R--symmetry based on the local N=2 $SU(2)$
gauge theory associated with the Calabi--Yau manifold
$WCP_4(8;2,2,2,1,1)$ of Hodge numbers $(h_{11}=2, h_{21}=86)$ that
has been considered in \cite{VafaKach} as an example of
heterotic/type II duality.

\par
As a final remark, we note that once the R-symmetry charges in the $N=2$ local
theory have been fixed, we may twist to the topological gravity theory
coupled to the axion-dilaton field the gauge fields and hypermultiplet scalars.
As it is well known, in the topological theory we have a set of instanton
conditions which fix the topological symmetry.
The  form of these instanton
conditions is universal and applies both to the classical and quantum case.
Specifically  it turns out that there are
four equations  describing the coupling of
four types of instantons:
\begin{equation}
\label{in1}
\begin{array}{rl}
\mbox{ i)} & \mbox{gravitational instanton}\\
\mbox{ ii)} & \mbox{gauge--instantons}\\
\mbox{ iii)} & \mbox{triholomorphic hyperinstantons}\\
\mbox{ iv)} & \mbox{H-monopoles}.
\end{array}
\end{equation}
Instanton equations of this type have already been discussed in
\cite{topftwist_1,topftwist_2,topf4d_8,topf4d_4};
the main difference is that in
\cite{topftwist_2,topf4d_8} the instanton conditions were only
the first three of eq.s (\ref{in1}).
The H-monopoles [8--12],
namely the instanton-like configurations
\begin{equation}
\label{in2}
\partial_{a} D\, =\,\epsilon_{abcd}\ee D H^{bcd}
\end{equation}
in the dilaton-axion sector were missing. In eq. (\ref{in2})
$D$ is the dilaton field and $H_{\mu\nu\rho}$ is the curl
of the antisymmetric axion
tensor $B_{\mu \nu}: \partial_{[\rho} B_{\mu\nu]}=H_{\mu\nu\rho}$.
The reason why they
were missing in
\cite{topftwist_2,topf4d_8} is the type of symmetry used there to define the
ghost number, namely an on-shell $R$-duality
based on the properties of the so-called minimal coupling.
The new type of gravitationally extended $R$-symmetry that we
present here is typically stringy in its origin and for
the classical moduli--spaces is an ordinary
off-shell
symmetry, which does not mix electric and magnetic states as the
$R$-duality
of the minimal case does. In the quantum--corrected effective
lagrangians $R$--symmetry reduces once again to an
$R$--duality, namely to a discrete
group of electric--magnetic duality rotations; yet the
preferred direction of the dilaton--axion field is
mantained also in the quantum case as it is necessary
on physical grounds. The new version of $R$--symmetry
discussed here
provides the solution to several conceptual problems at the same time.

More details on the topics discussed in this contribution can be found
in \cite{rsym} and references therein.

\end{document}